\documentclass[aps,prb,twocolumn,amsmath,amssymb,superscriptaddress,longbibliography]{revtex4-1}
\usepackage[dvips]{graphicx}
\usepackage[utf8]{inputenc}
\usepackage{dcolumn}
\usepackage{physics}
\usepackage{bm}
\usepackage{graphics}
\usepackage{dirtytalk}
\usepackage{epsfig,color}
\usepackage[breaklinks,colorlinks,bookmarks=false,citecolor=red,linkcolor=blue,urlcolor=magenta]{hyperref}
\graphicspath{{Fig/}} 

\begin{document}
\title[J. Vahedi ]{Entanglement and quantum correlations in
the XX spin-$1/2$ honeycomb lattice}
\author{S. Satoori}
\address{Department of Physics, University of Guilan, 41335-1914, Rasht, Iran.} 

\author{S. Mahdavifar}
\email{smahdavifar@gmail.com}
\address{Department of Physics, University of Guilan, 41335-1914, Rasht, Iran.} 

\author{J. Vahedi}
\email{jvahedi@jacobs-university.de}
\affiliation{Department of Physics and Earth Sciences, Jacobs University Bremen, Bremen 28759, Germany.}

\date{\today}
\begin{abstract}
The ground state phase diagram of the dimerized spin-1/2 XX honeycomb model in presence of a transverse  magnetic field (TF) is known. With the absence of the magnetic field, two quantum phases, namely, the N\'eel and the dimerized phases have been identified. Moreover, canted N\'eel and the paramagnetic (PM) phases also emerge by applying the magnetic field. In this paper, using two complementary numerical exact techniques, Lanczos exact diagonalization, and Density matrix renormalization group (DMRG) methods, we study this model by focusing on the quantum correlations, the concurrence, and the quantum discord (QD) among nearest-neighbor spins. We show that the quantum correlations can capture the position of the quantum critical points in the whole range of the ground state phase diagram consistent with previous results. Although the concurrence and the QD are short-range, informative about long-ranged critical correlations. In addition, we address a “magnetic-entanglement” behavior that starts from an entangled field around the saturation field.
\end{abstract}

\maketitle
\section{Introduction}\label{sec1}
The dimerization phenomenon can emerge at zero-temperature behavior of low-dimensional spin-$1/2$ systems. Interactions favor the spin-singlet (or triplet) between pair of spins, and therefore the ground state is a superposition of dimer states. The quantum dimer systems were initially  proposed as a mapping of the lattice Bose gas to the quantum antiferromagnets~\cite{R1}.
\par
In the past two decades, searching for spin-1/2 dimerized honeycomb structures have attracted much interest from an experimental point of view ~\cite{R2,R3,R4,R5,R6,R7,R8,R9,R10,R11,R12}. Many materials have been realized as dimerized spin-1/2 honeycomb antiferromagnets. For example, Cu${}_{2}{A}_{2}$O${}_{7}$ is known as a distorted honeycomb lattice~\cite{R8}. A phase transition to an antiferromagnetically ordered state at $0.77K$ is reported for the Verdazyle radical $2\text{\ensuremath{-}}\mathrm{Cl}\text{\ensuremath{-}}3,6\text{\ensuremath{-}}{\mathrm{F}}_{2}\text{\ensuremath{-}}\mathrm{V}$~\cite{R9}.  In addition, no long-range magnetic order is observed down to $0.6K$ in the specific heat measurements of a polycrystalline sample of the spin-1/2 distorted honeycomb lattice antiferromagnetic Cu${}_{2}{A}_{2}$O${}_{7}$~\cite{R10}. In very recent work, it is shown that two antiferromagnetic interactions lead to the formation of a honeycomb lattice in some  verdazyl-based complexes~\cite{R12}.        
\par
Theoretically, the effect of dimerization on the physics of spin-1/2 honeycomb lattices  was the subject of many studies.  In absence of the dimerization, it is known to realize N\'eel long-range order phase at zero temperature~\cite{R14,R15,R17,R18,R19}. In the presence of dimerization, transforming the spin system onto a nonlinear sigma model, the ground state phase diagram consisting N\'eel and disordered spin gap phases has been proposed~\cite{R20}. The mentioned quantum phase transition is confirmed by  numerical quantum Monte Carlo ~\cite{R21} and tensor renormalization-group method ~\cite{R22}. By presenting the randomness on the exchange interaction in a spin-1/2 honeycomb lattice, a quantum spin liquid phase appear in the ground state phase diagram~\cite{R23,R24}. By doing triplon analysis and quantum Monte Carlo calculations, a spin-1/2 Heisenberg model on the honeycomb lattice with three different antiferromagnetic exchange interactions is also studied~\cite{R25}. The existence of plateau states are reported in the magnetization process in this model. Also, the spin-1/2 dimerized model on a honeycomb lattice with antiferromagnetic and ferromagnetic  interactions is systematically studied using the continuous-time quantum Monte Carlo method~\cite{R26,R27,R28}.
\par
In recent years,  powerful approaches based on the concepts borrowed from the quantum information theory~\cite{R29} have been developed and intensively used to identify  quantum critical points in different complex many-body systems~\cite{R30}. In particular, the detailed analysis of various bipartite quantum correlations as the entanglement and the QD, has been successfully exploited to tackle many complicated problems~\cite{R31,R32,R33,R34,R35,R36,R37,R38,R39,R40,R41,R42,R43,R44,R45,R46,R47,R51,R52,R53,R55,R56,R57,R58,R59}.
\par
\begin{figure}[t]
 \includegraphics[width=1\columnwidth]{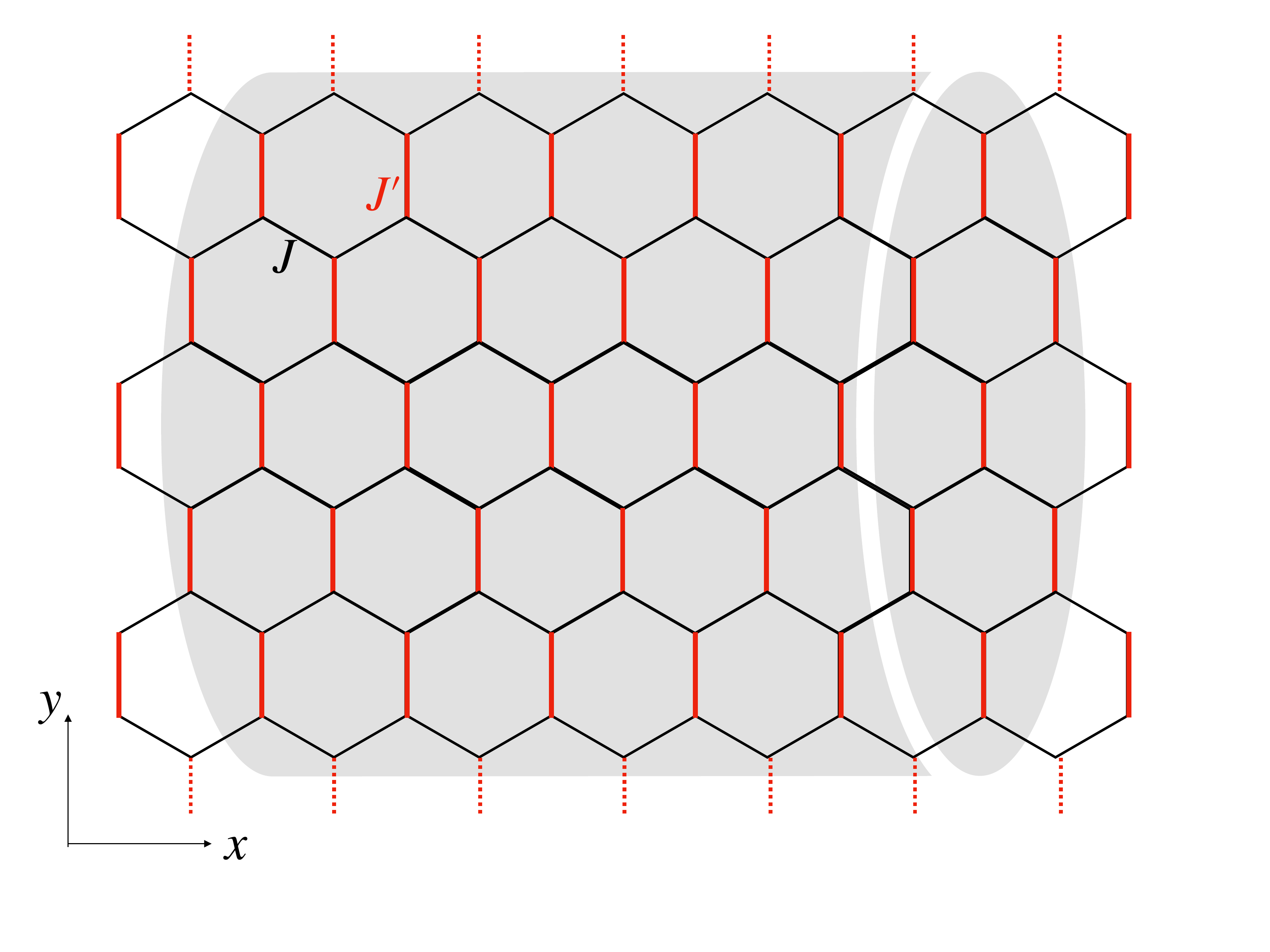}\\
 \includegraphics[width=1\columnwidth]{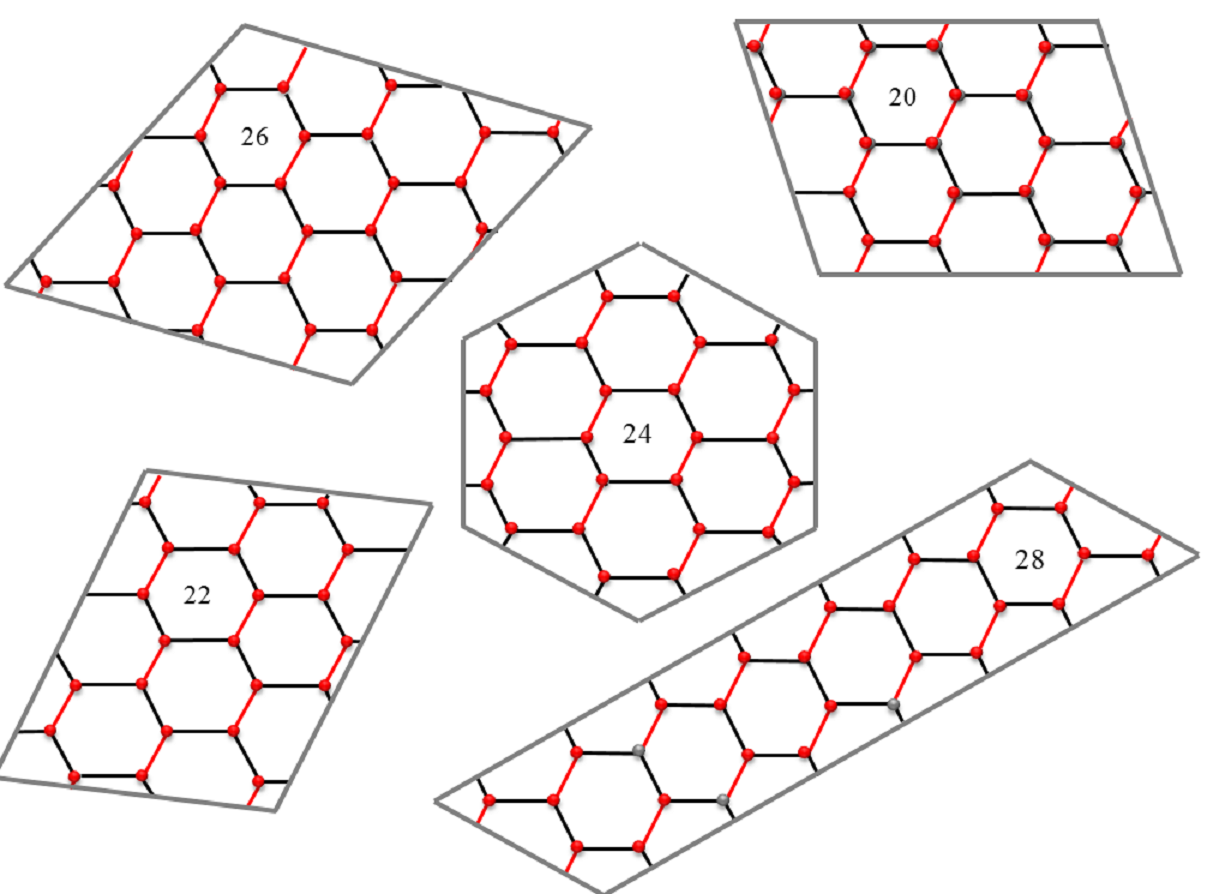}
\caption{(colour online) Schematic picture of the honeycomb lattice with different antiferromagnetic interaction coupling $J$ and $J'$, as shown in black and red lines, respectively. The top and bottom panels show cylindrical and flak finite-size clusters considered within the numerical Lanczos and DMRG methods. For the cylindrical clusters, a periodic boundary is applied in the $y$-diresction, while for the flak shapes a twist periodic boundary is considered.}
\label{fig1}
\end{figure}

Motivated by this, we study the 2D dimerized spin-1/2 XX honeycomb model in the presence of a TF. We have used the exact numerical Lanczos and DMRG techniques to probe entanglement features with dimerizatoin parameter. The dimerization parameter is defined as $\alpha={J}/{J'}$ (shown on Fig.~\ref{fig1})). Two kinds of pairs can be considered: (1) on a bond with coupling $J$, (2) on a bond with  coupling $J'$. Our numerical results show that entanglement between pairs of spins on $J$-bonds, signaled the quantum critical point between the  N\'eel and the dimerized phases. By applying the TF,  a magnetic entanglement is recognized that starts from a critical entangled field around the saturation field. In addition, all ground-state phases have discussed from the viewpoint of quantum correlations.
\par
The rest of the paper is organized as follows. In the next section, the model is introduced. In section III, a short review of quantum correlations as the entanglement and the QD are given. Section IV presents numerical Lanczos and DMRG results on finite-size clusters. Finally, in Section V, we summarize our conclusion.

\section{Model} \label{sec2}
In this section, we consider the antiferromagnetic dimerized XX model on the honeycomb lattice. The Hamiltonian is defined  as
\begin{eqnarray}
H&=&J \sum_{<i,j>}(S_i^xS_j^x+S_i^yS_j^y) +J'\sum_{<i,j>'} (S_i^xS_j^x+S_i^yS_j^y)\nonumber\\
 &-&h\sum_{i=1}S_i^z ,
 \label{eq1}
\end{eqnarray}
where $S_i$ is the spin-$\frac{1}{2}$ operator on the $i$-th site of the lattice. $<i,j>$ and $<i,j>'$, with different antiferromanetic interaction exchange  couplings $J$ and $J'$ respectively, run over all the nearest neighbours (as schematic picture in Fig.~\ref{fig1}). $h$ denotes the TF. In absence of the TF, $h=0$, a critical dimerization value $\alpha_c$ which separates the N\'eel and the dimerizad phases. At  region with $\alpha<\alpha_c$,  a phase transition into the paramagnetic (PM) phase anticipate occurs  at the critical saturation field $h=h_s(\alpha)$. However, in the dimerized phase, two quantum phase transitions have  been reported\cite{R22}. First, model undergoes a phase transition from the dimerized into the canted N\'eel phase at $h=h_{c_{1}}(\alpha)$. Second, by more increasing the TF, system goes to the PM phase at $h=h_s(\alpha)$.  
\par
The theoretical quantum study of such a physical problem requires appropriate handling of very high-rank matrices. Although the matrix of the Hamiltonian is sparse, using the standard methods it is not possible to solve the problem by direct diagonalization of a very large matrix. In the following, we apply two of the most impressive numerical tools, called the numerical Lanczos and DMRG methods for computing ground state of the Hamiltonian and then extract quantum correlations on finite size systems. The numerical Lanczos method with appropriate implementations has emerged as one
of the most applicable computational procedures, mainly 
when the ground state is desired~\cite{R60}. 
\par
Although the numerical Lanczos technique allows for the exact analyses of the model's ground state, the disadvantage is, of course, its limitation to small system sizes. To study bigger system sizes,  one idea is the matrix-product state (MPS) based methods, such as density matrix renormalization group (DMRG)\cite{dmrg1992,dmrg2005}. The DMRG gives access to the ground state wavefunction from which one can compute observables. The DMRG calculations in this paper performed using the ITensor C++ library (version 3.1)~\cite{itensor}. We run sweeps for the entropy to converge to at least $10^{-10}$, and a large number of states, up to 1000, was kept so that the truncation error is less than $10^{-12}$.
\section{Quantum Correlation}\label{sec3}
Quantum correlations have become central for the characterization and classification of many-body quantum systems. Peculiar zero-temperature quantum phases such as spin liquids~\cite{R61,R62}, topological~\cite{R63,R64,R65}, and many-body localized systems ~\cite{R66,R67,R68} find their hallmarks in their quantum correlation features. It should be noted  that the entanglement in many-body systems can be accessible in experiments such as in full-state tomography~\cite{R69,R70}  and  ultra-cold atoms to measure Renyi entropies~\cite{R71,R72}. Besides, quantum phase transitions are signaled by a universal quantum correlation contribution determined solely by the universality class of the quantum phase transitions~\cite{R73,R74,R75,R76,R77,R78}. Hence, they can be used to  detect quantum phase transitions without prior knowledge of the nature of the transition.
\par
For a pair of spin-1/2 particles, it has been shown that the concurrence which is essentially equivalent to the entanglement of formation, can be taken as a measure of entanglement. The concurrence between two spins at sites $i$ and $j$ is determined by the corresponding reduced density matrix $\rho_{ij}$,
\begin{eqnarray}
\rho_{ij}=\left(
\begin{array}{cccc}
X_{ij}^{+} & 0 & 0 & 0 \\
0 & Y_{ij}^{+} & Z_{ij}^{*} & 0 \\
0 & Z_{ij} & Y_{ij}^{-} & 0 \\
0 & 0 & 0 & X_{ij}^{-} \\
\end{array}
\right),
\label{density matrix2}
\end{eqnarray}
where non-zero elements of the density matrix are given by 
\begin{eqnarray} 
X_{ij}^{+}&=& \langle(1/2+S_{i}^{z})(1/2+S_{j}^{z})\rangle,\nonumber\\
Y_{ij}^{+}&=& \langle(1/2+S_{i}^{z})(1/2-S_{j}^{z})\rangle,\nonumber\\
Y_{ij}^{-}&=& \langle(1/2-S_{i}^{z})(1/2+S_{j}^{z})\rangle,\\
X_{ij}^{-}&=& \langle(1/2-S_{i}^{z})(1/2-S_{j}^{z})\rangle,\nonumber\\
Z_{ij}&=& \langle S_i^{+}S_{j}^{-}\rangle.
\label{dm2}
\end{eqnarray}
The concurrence is obtained by the following expression:
\begin{eqnarray} 
C_{ij} =2 \max{\{0, |Z_{ij}|-\sqrt{X_{ij}^{+}X_{ij}^{-}}\}}.
\label{Concurr0}
\end{eqnarray}
One should notes that, there are different quantum correlations that are not spotlighted by the entanglement measures.  These quantum correlations are thoroughly included in the formulation of so-called the QD as a measure for representing all quantum correlations~\cite{R79,R80}. It is defined as  the difference between the mutual information, ${\cal I}({\rho _{ij}})$, and classical correlations ${\cal C}({\rho _{ij}})$: 

\begin{equation}
QD_{ij} = {\cal I}({\rho _{ij}}) - {\cal C}({\rho _{ij}}).
\end{equation}
Mutual information does a measure on the correlation between pair spins $S_i$ and $S_j$ and is given by
\begin{equation}
{\cal I}({\rho _{ij}}) = S({\rho _{i}}) + S({\rho _{j}}) + \sum\limits_{\alpha  = 0}^3 {{\lambda _\alpha }} \log ({\lambda _\alpha }),
\end{equation}
where ${\lambda _{\alpha}} $ are eigenvalues of the reduced density matrix, $\rho _{ij}$. By definition new variables
\begin{equation}
\begin{array}{l}
{c_1} = 2{Z_{ij}},\\
{c_2} = X_{ij}^ +  + X_{ij}^ -  - Y_{ij}^ +  - Y_{ij}^ - ,\\
{c_3} = X_{ij}^ +  - X_{ij}^ - ,
\end{array}
\end{equation}
the entropy is determined as
\begin{equation}
\begin{array}{l}
S({\rho _i}) = S({\rho _{j}}) = \\
\qquad\quad - \left[ {(\frac{{1 + {c_3}}}{2})\log (\frac{{1 + {c_3}}}{2}) + (\frac{{1 - {c_3}}}{2})\log (\frac{{1 - {c_3}}}{2})} \right].
\end{array}
\end{equation}

On the other hand, by definition 
\begin{equation}
\begin{array}{l}
{q_{k1}} = {( - 1)^k}{c_1}\left[ {\frac{{\sin (\theta )\cos (\phi )}}{{1 + {{( - 1)}^k}{c_3}\cos (\theta )}}} \right],\\
{q_{k2}} = \tan (\phi ){q_{k1}},\\
{q_{k3}} = {( - 1)^k}\left[ {\frac{{{c_2}\cos (\theta ) + {{( - 1)}^k}{c_3}}}{{1 + {{( - 1)}^k}{c_3}\cos (\theta )}}} \right],\\
{\theta _k} = \sqrt {{q_{k1}^2}+{q_{k2}^2}+{q_{k3}^2} }
\end{array}
\end{equation}
where $0 \le \theta  \le \pi $,  $0 \le \phi  \le 2\pi $ and $k=0, 1$. The classical correlations, ${\cal C}({\rho _{ij}})$ can be obtained by

\begin{equation}
\begin{array}{l}
{\cal C}({\rho _{ij}}) = \\
\mathop {\max }\limits_{\left\{ {\prod\nolimits_i^B } \right\}} \left( {S({\rho _i}) - \frac{{S({\rho _0}) + S({\rho _1})}}{2} - {c_3}\cos (\theta )\frac{{S({\rho _0}) - S({\rho_1})}}{2}} \right), 
\end{array}
\end{equation}
where 
\begin{equation}
\begin{array}{l}
S({\rho_k}) =  - \left( {\frac{{1 + {\theta_k}}}{2}} \right)\log \left( {\frac{{1 + {\theta _k}}}{2}} \right) \\
\qquad\qquad + \left( {\frac{{1-{\theta_k}}}{2}} \right)\log \left( {\frac{{1-{\theta_k}}}{2}} \right).
\end{array}
\end{equation}

\section{Numerical results} \label{sec4}
Here, we present the numerical results based on the Lanczos and DMRG methods. Twist periodic boundary condition (PBC) is applied for honeycomb lattice with finite flake sizes  $N=20, 22, 24, 26, 28$ for both the Lanczos and DMRG methods. Moreover, we consider cylinder clusters in the DMRG method with PBC in the $y$-direction (as shown in Fig.~\ref{fig1}).  Having the ground state of the system, $|GS\rangle$, then quantum correlations as the concurrence and the QD are obtained.
\par
\begin{figure}[t]
\centerline{\psfig{file=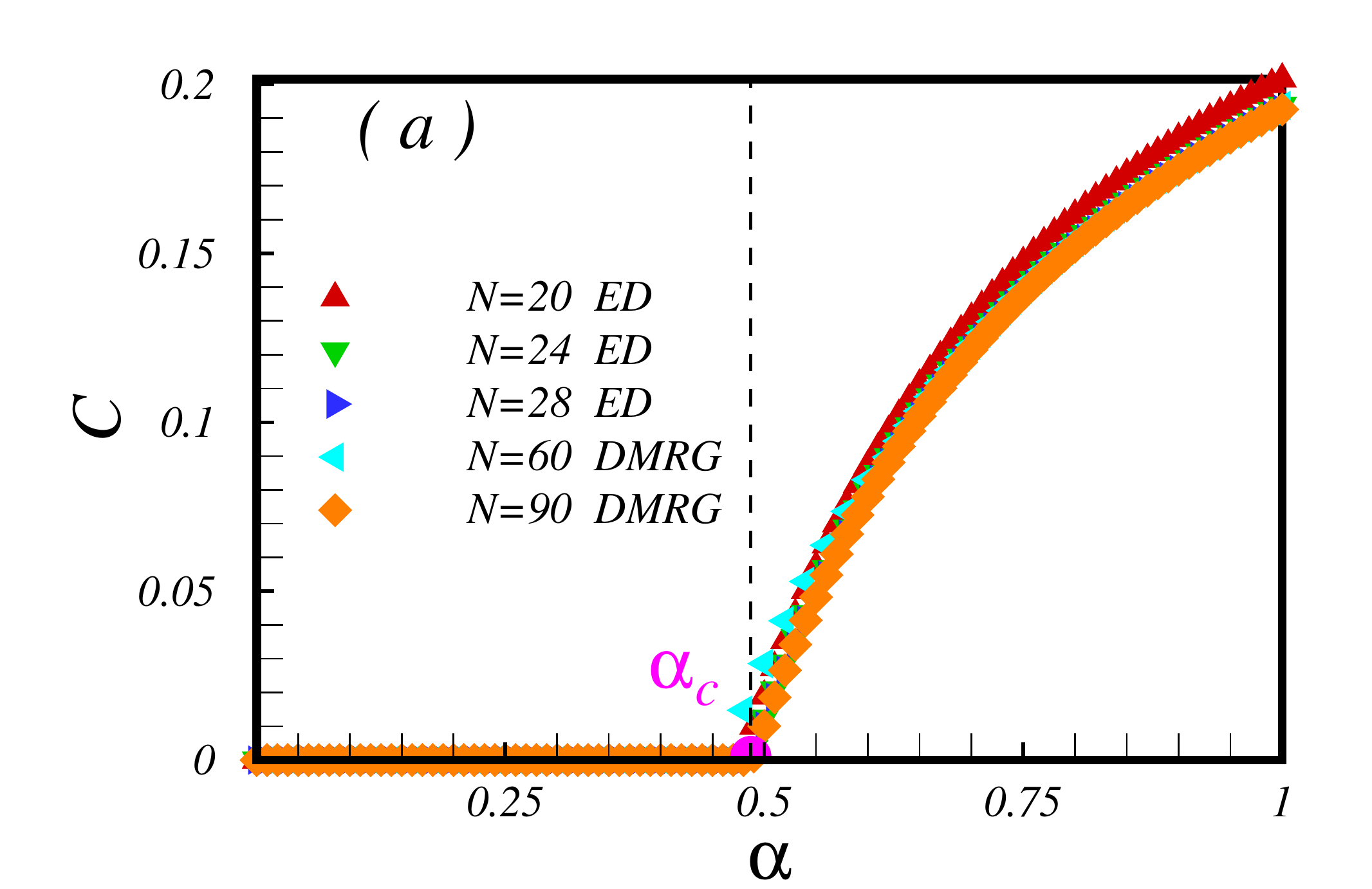,width=1.8in} \psfig{file=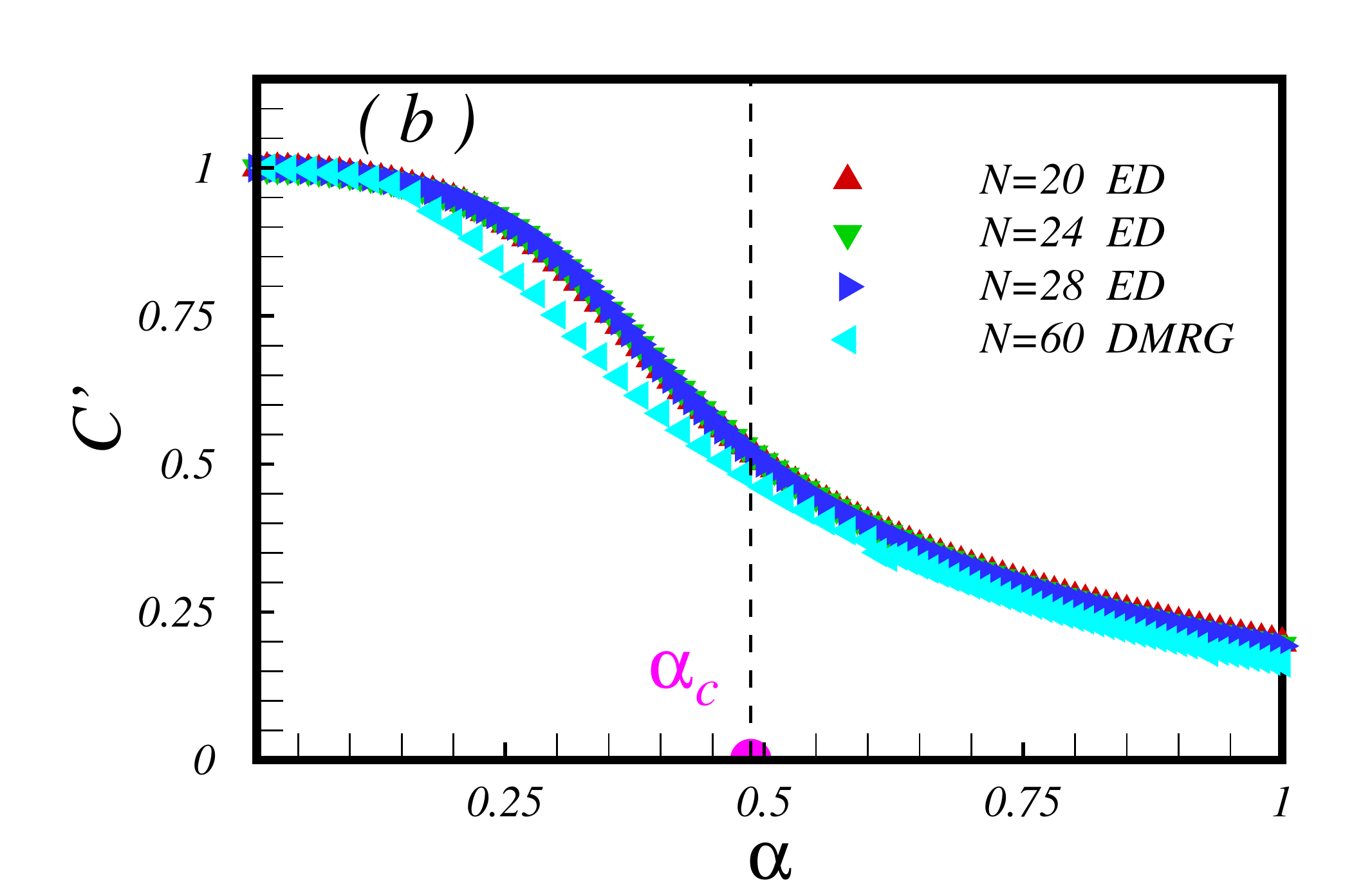,width=1.8in}}
\centerline{\psfig{file=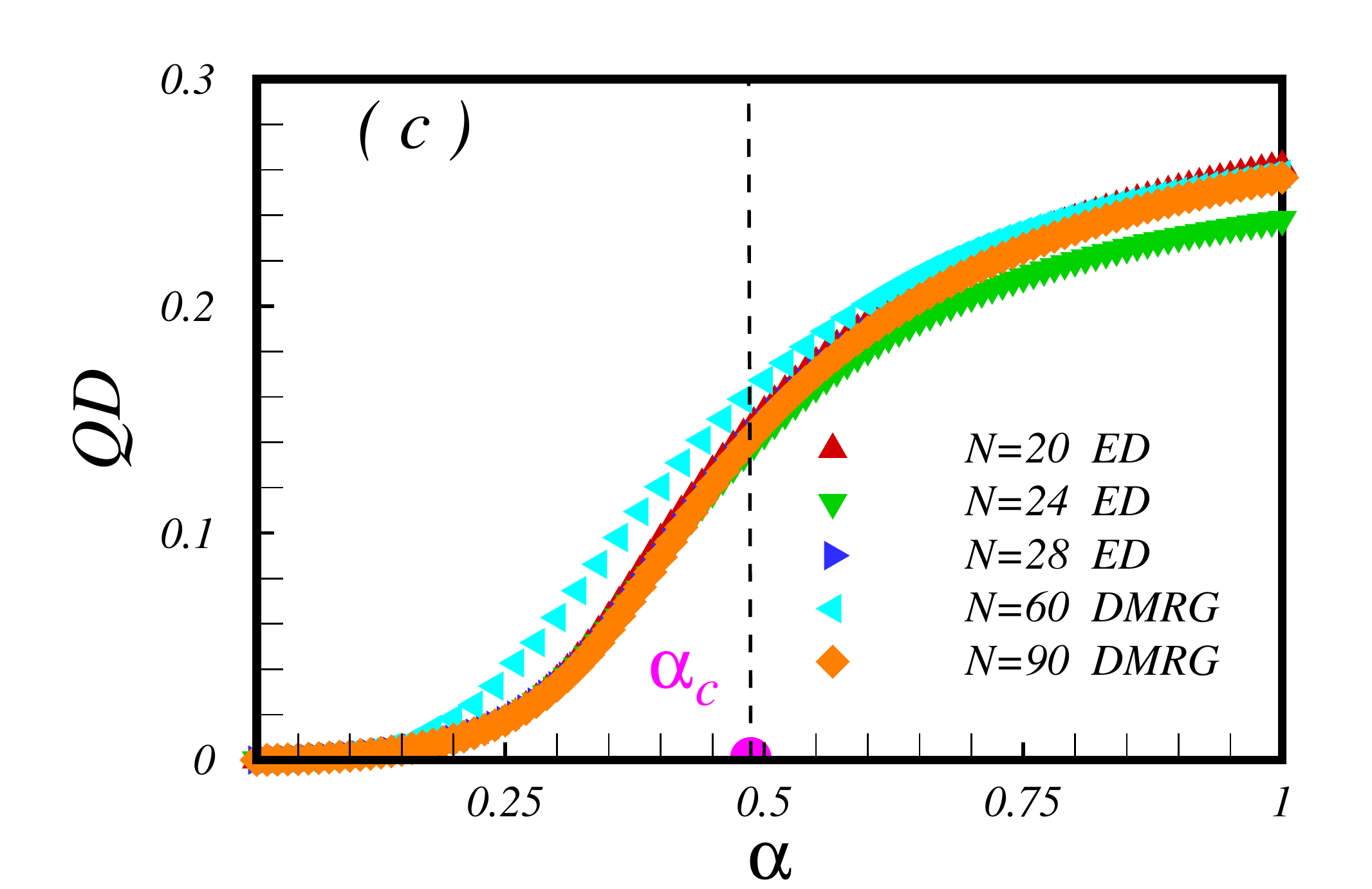,width=1.8in} \psfig{file=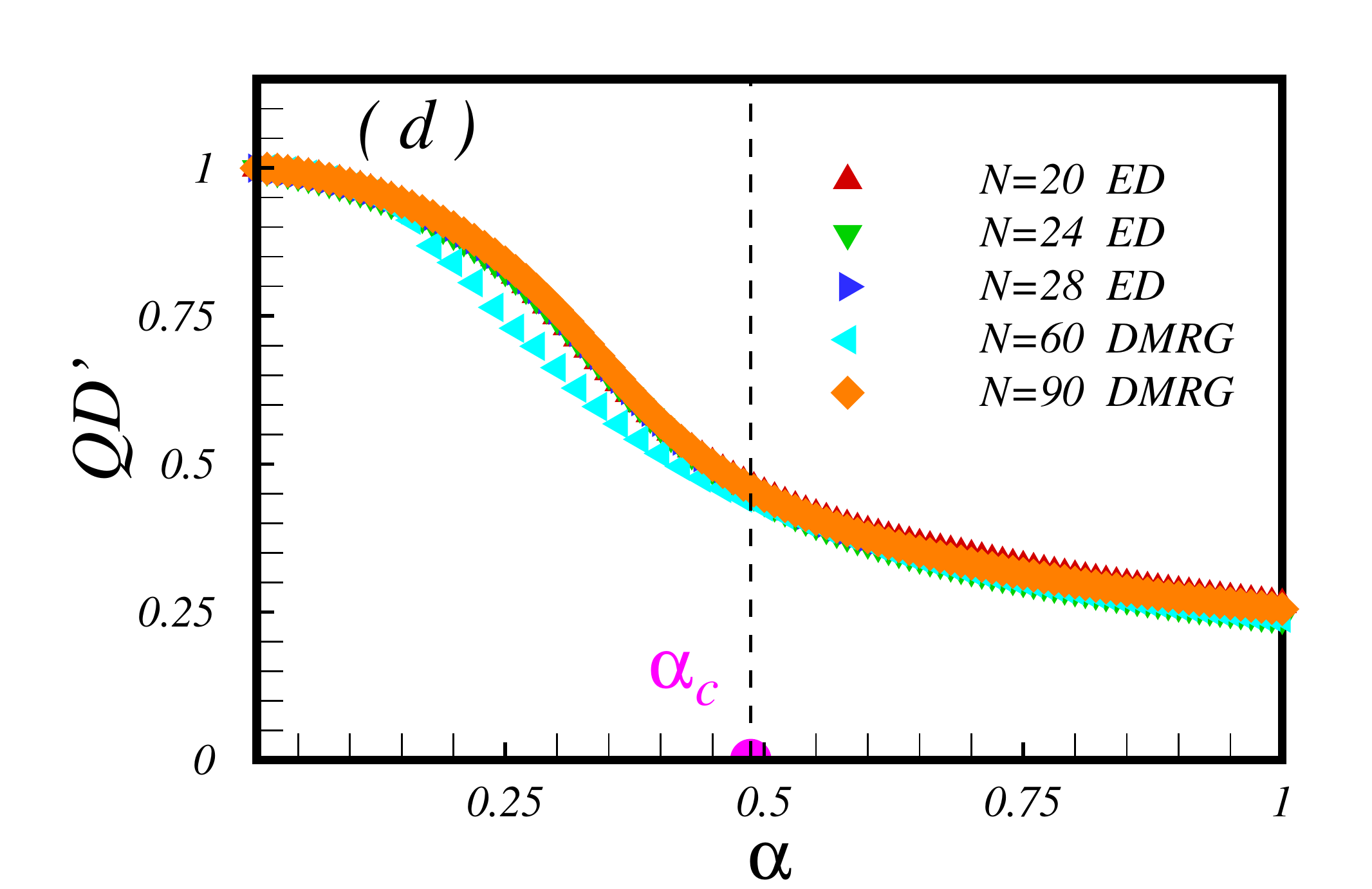,width=1.8in}}
\centerline{\psfig{file=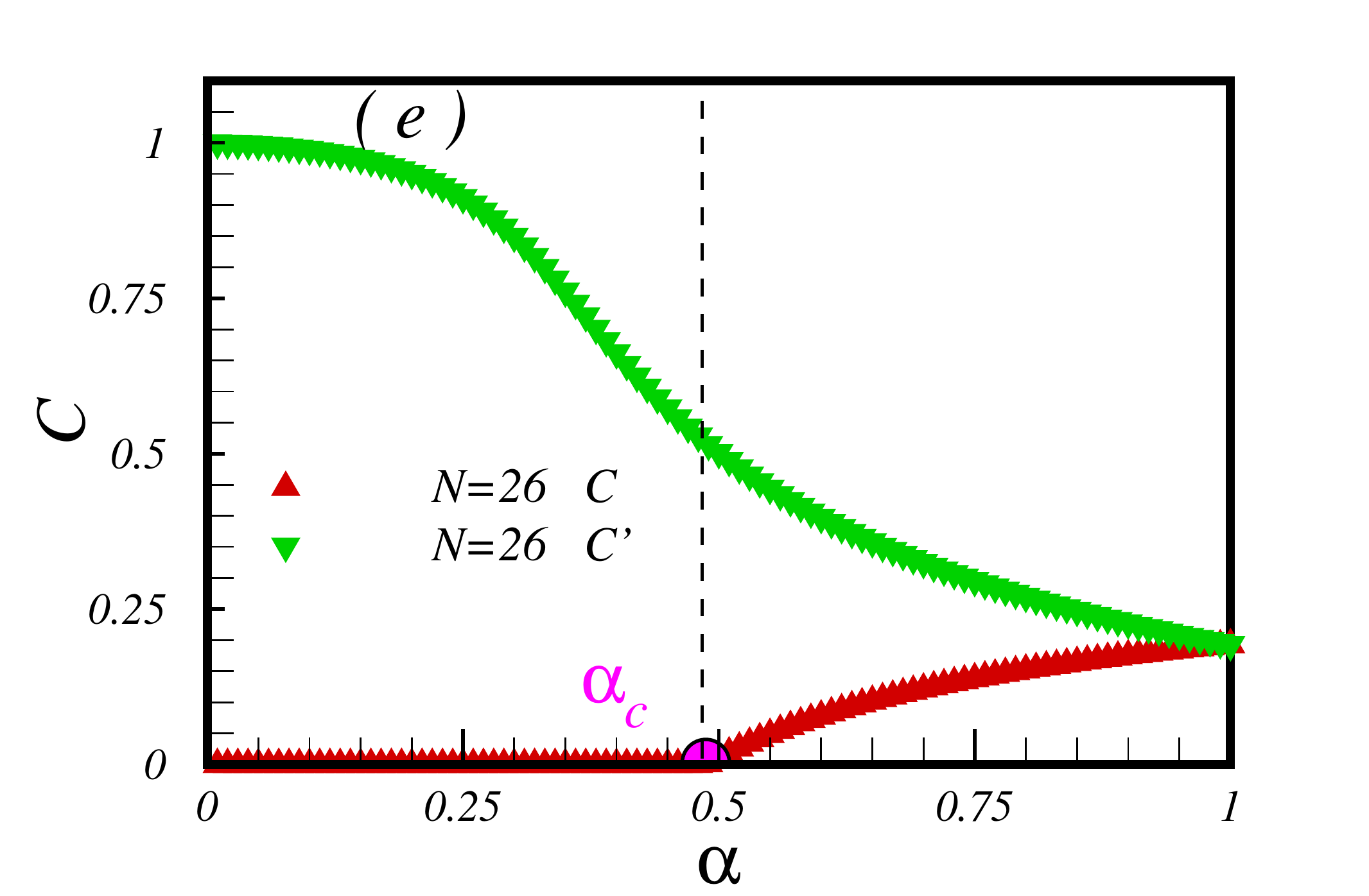,width=1.8in} \psfig{file=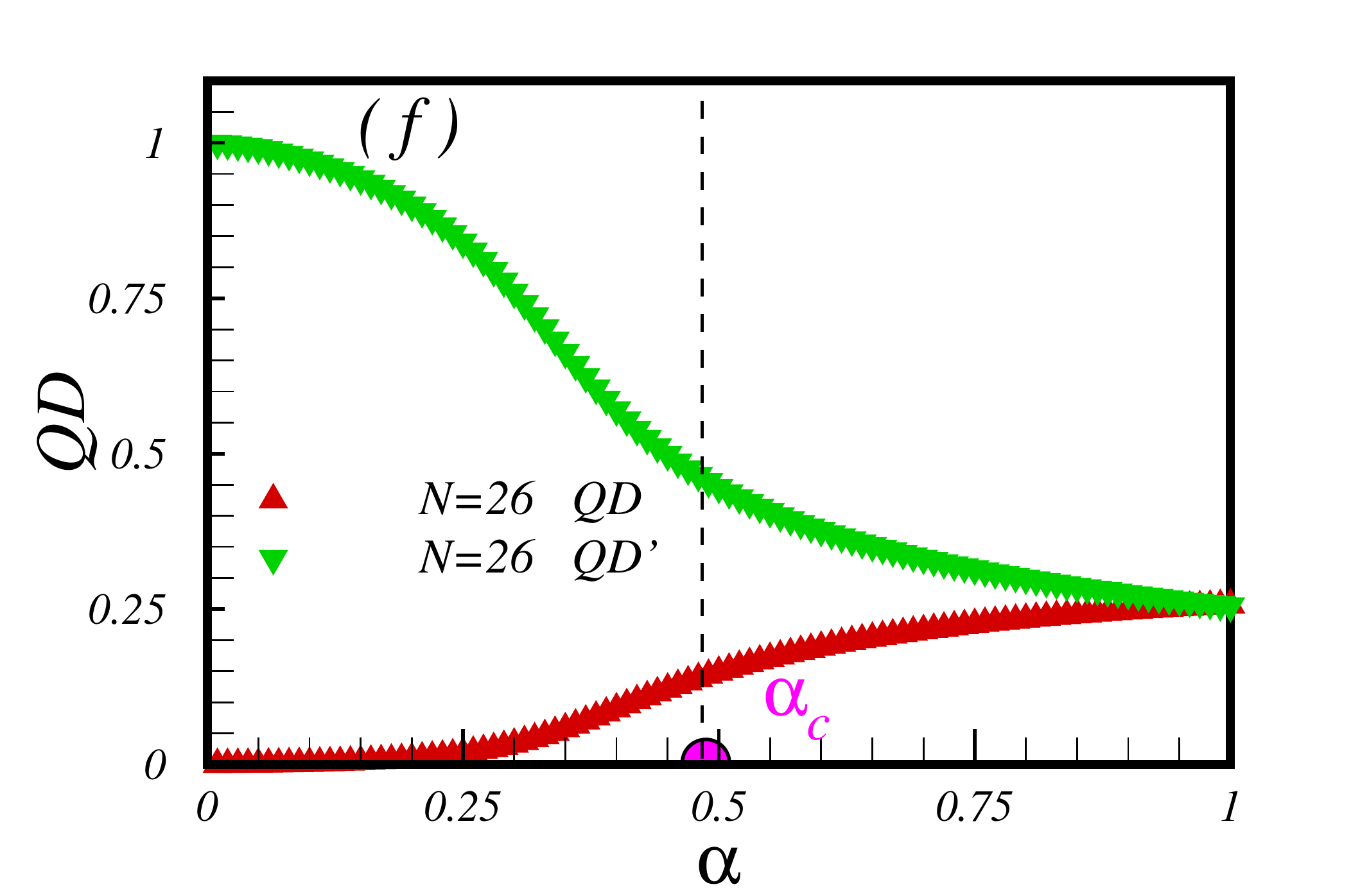,width=1.8in}}
\caption{(color online) The concurrence and the QD between pair of spins on bonds with exchange coupling $J$ ((a) and (c)) and $J'$ ((b) and (d)). Lanczos results are presented for clusters with $N=20, 24, 28$ spins and also DMRG results for $N=60, 90$. In panels (e) and (f), concurrence and QD are plotted for a cluster with $N=26$ spins. At $\alpha=1$, no difference between concurrences (or values of QDs) is observed.}
\label{QC}
\end{figure}

First, we consider the model in the absence of a magnetic field.  In Fig.~\ref{QC}, the numerical results of the concurrence and the QD between pair of spins on a bond with exchange coupling $J$ $(C)$ and on a bond with exchange coupling $J'$ $(C')$ are presented. In the case, $\alpha=0$, the honeycomb system divides into $N/2$ individual pair spins where at zero temperature are in the singlet state (are also called dimers). Pair spins in the singlet state are maximally entangled.  Consistent with this picture, numerical results in  Fig.~\ref{QC} (a)  and (b) show that, at $\alpha=0$, only pair spins on bonds with exchange coupling $J'$ are maximally entangled and others with exchange coupling $J$ are unentangled. Now by turning $J$, what we found is interesting, the model still can be effectively treated as dimers  (see panel Fig.~\ref{QC}-(a)). That is almost true up a critical point, namely $\alpha_c$, which concurrence remains zero on bonds with exchange coupling ${J}$. As soon as the dimerization parameter increases from $\alpha_c$, where the model goes into the N\'eel phase, pair spins on bonds with exchange coupling $J$ entangled and signature of the mentioned critical point is clearly observe in the behavior of $C$ (see panel Fig.~\ref{QC}-(a)). On the other hand, in the limit  $\alpha \longrightarrow \infty$, the honeycomb system divides into individual spin-1/2 XX chains. The ground state of an individual chain system is in the Luttinger-liquid phase and it is known that the nearest neighbours are entangled ~\cite{R30,R31,R44,R45}. Consistent with this picture, our numerical results show that, only pair spins on bonds with exchange coupling $J$ are entangled with $C \simeq 0.34$. Critical dimerization  $\alpha_c = 0.48 \pm 0.02$  and $\alpha_c = 0.5 \pm 0.02$  are found within  the Lanczos and the DMRG, respectively. The difference could pertain to the finite size effect and different clusters used on the two approaches.
\par
In addition to the concurrence, results of the QD are  plotted in Figs.~\ref{QC} (c)  and (d). In the case, $\alpha=0$, QD exists only between pair of spins on dimers. Interestingly, by switching $\alpha$ on, QD as quantum correlations, but not necessarily involve quantum entanglement, developed between spins on bonds with ${J}$. As can be seen, QD between pair of spins on  bonds with exchange coupling ${J'}$ show decreasing behavior in contrast with those on bonds with exchange coupling ${J}$. Though the finite QD is an indication of a reach ground state for $0<\alpha<\alpha_c$, it is not showing any signature as passing the quantum critical point. In the limit  $\alpha \longrightarrow \infty$, where model divides into individual spin-1/2 XX chains, we found that our numerical results are in agreement with results obtained on a spin-$1/2$ chain model~\cite{R45,R81}.
\par
For the comparison purpose, in Figs.~\ref{QC} (e) and (f)   concurrence and QD on different bonds are depicted. As is observed, at $\alpha=1$ where the model becomes uniform,  either concurrence or QD   on different bonds cross each other. At this point, spins at two sublattices are aligned in an opposite direction to minimize the energy. It believes the model shows N\'eel order at zero temperature\cite{R14,R18}.
\begin{figure}[t]
\centerline{\psfig{file=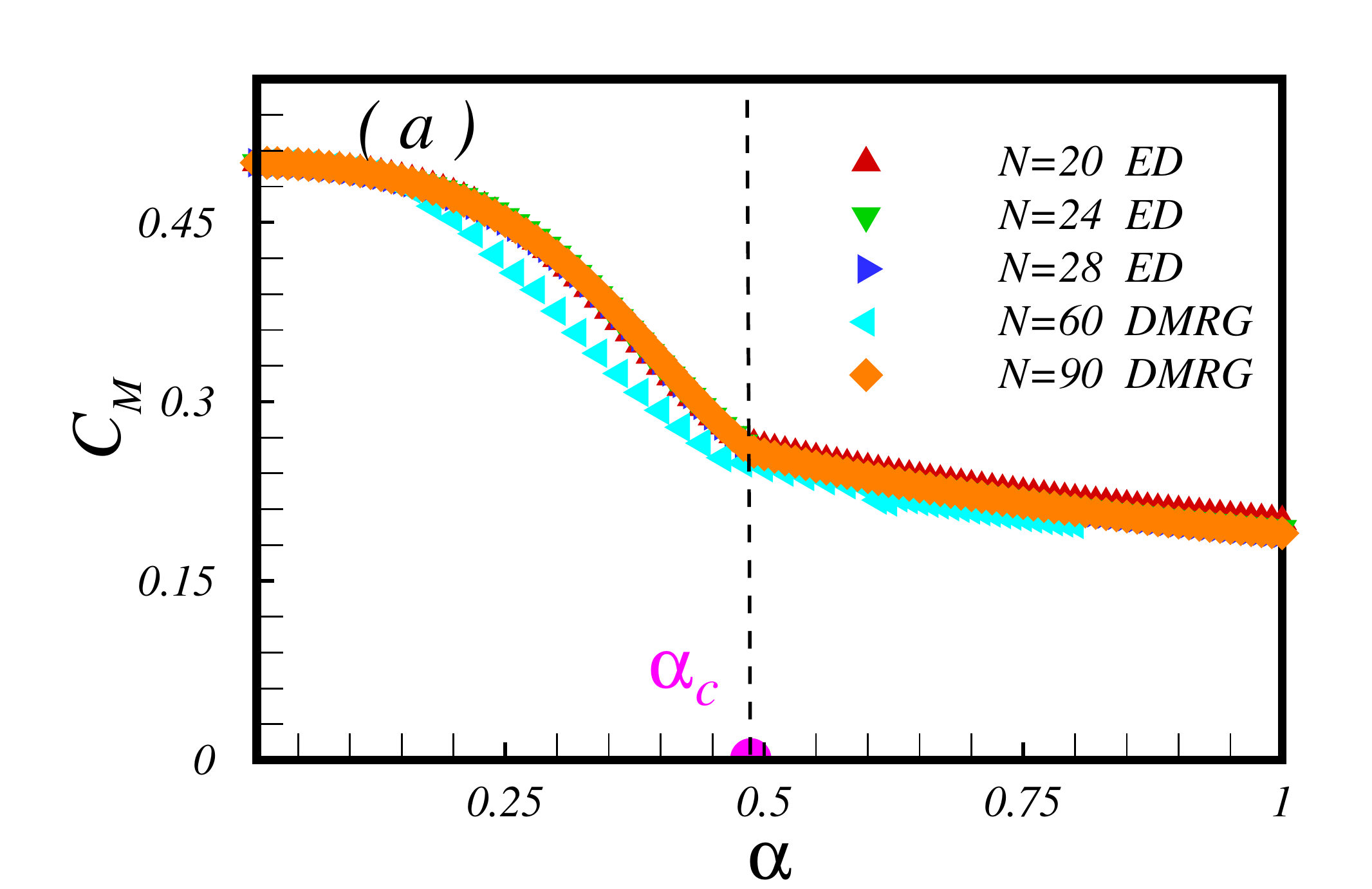,width=1.8in} \psfig{file=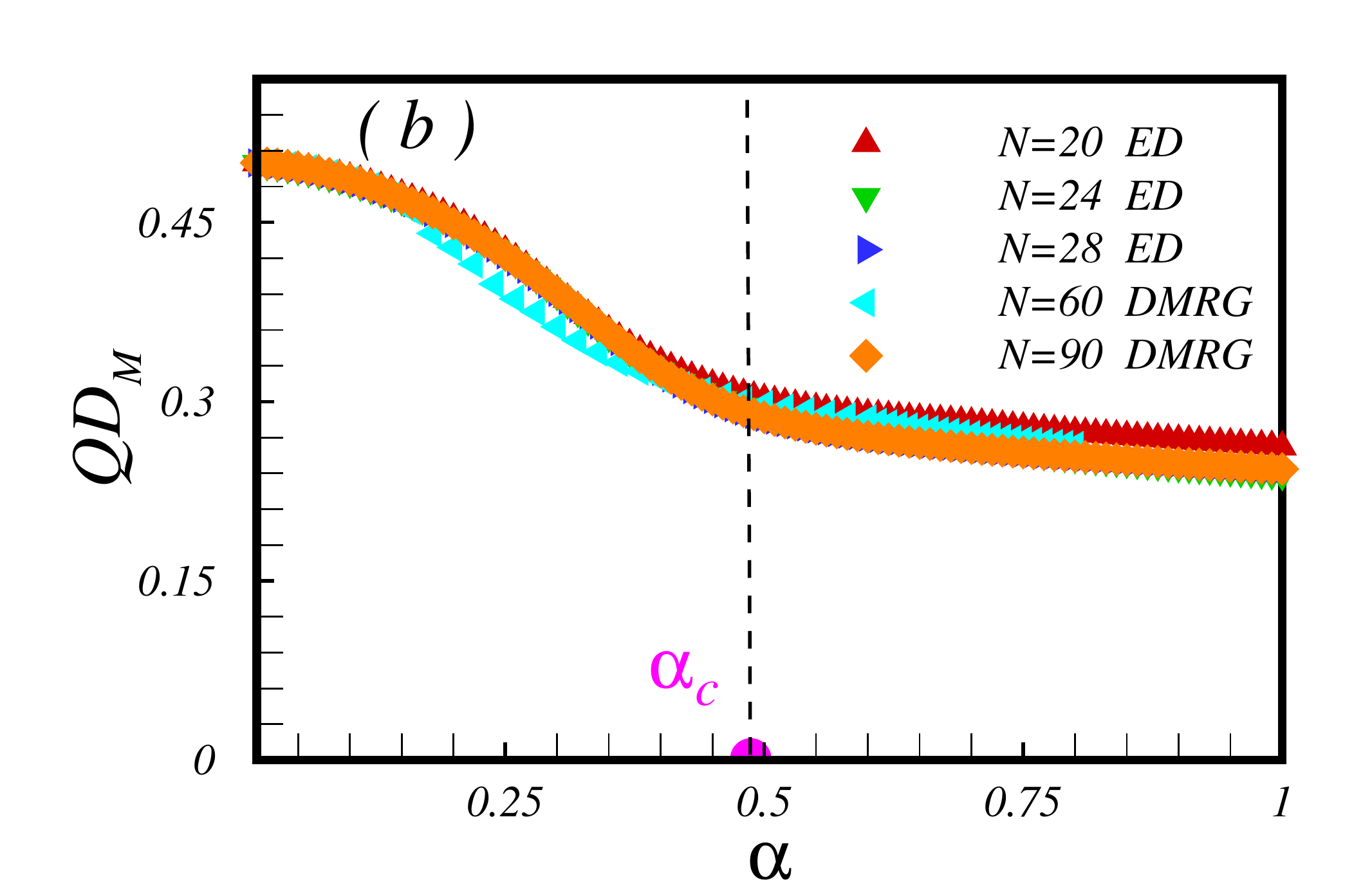,width=1.8in}}
\caption{(color online) Mean value of the (a) the concurrence and (b) the QD versus the dimerization parameter. Signature of the quantum critical point is clearly seen in the behaviour of the concurrence.}
\label{CM}
\end{figure}
\par
Within both numerical approaches, the ED and the DMRG, we  probe all pairs of spins. Then we introduce a mean measurement of the  concurrence and QD throughout the lattice as follows,
\begin{eqnarray} 
C_{{\rm {M}}} &=&\frac{1}{N'} \sum_{<i,j>}C_{ij}, \nonumber \\
QD_{{\rm {M}}} &=&\frac{1}{N'} \sum_{<i,j>}QD_{ij},
\label{Concurr_}
\end{eqnarray}
where $N'=\frac{3}{2} N$ is the number of pair spins in each cluster of the model. Results illustrate in  Fig.~\ref{CM}. As can be seen, $C_{{\rm M}}$ first decreases by increasing the dimerization parameter up to the quantum critical point $\alpha=\alpha_c$. As already seen, up to the critical point $\alpha_c$, all entanglement contributions to $C_{{\rm M}}$ come from bonds with exchange coupling ${J'}$ (bounds depicted with red color in Fig.\ref{fig1}). For dimerization parameter bigger than $\alpha_c$ spins between bonds with exchange coupling ${J}$ begin to entangle, and $C_{{\rm M}}$  shows almost a different decreasing slope in the region $\alpha>\alpha_c$. Thus the quantum critical point may be detected by focusing on the mean value of entanglement between pair of spins. However, the mean value of the QD between the nearest-neighbour pair of spins do not show the quantum critical point, as shown in Fig.~\ref{CM} (b).  Indeed, for the present model, by focusing only on the $QD_{{\rm M}}$, one could not detect the quantum critical point.

\begin{figure}[t]
\centerline{\psfig{file=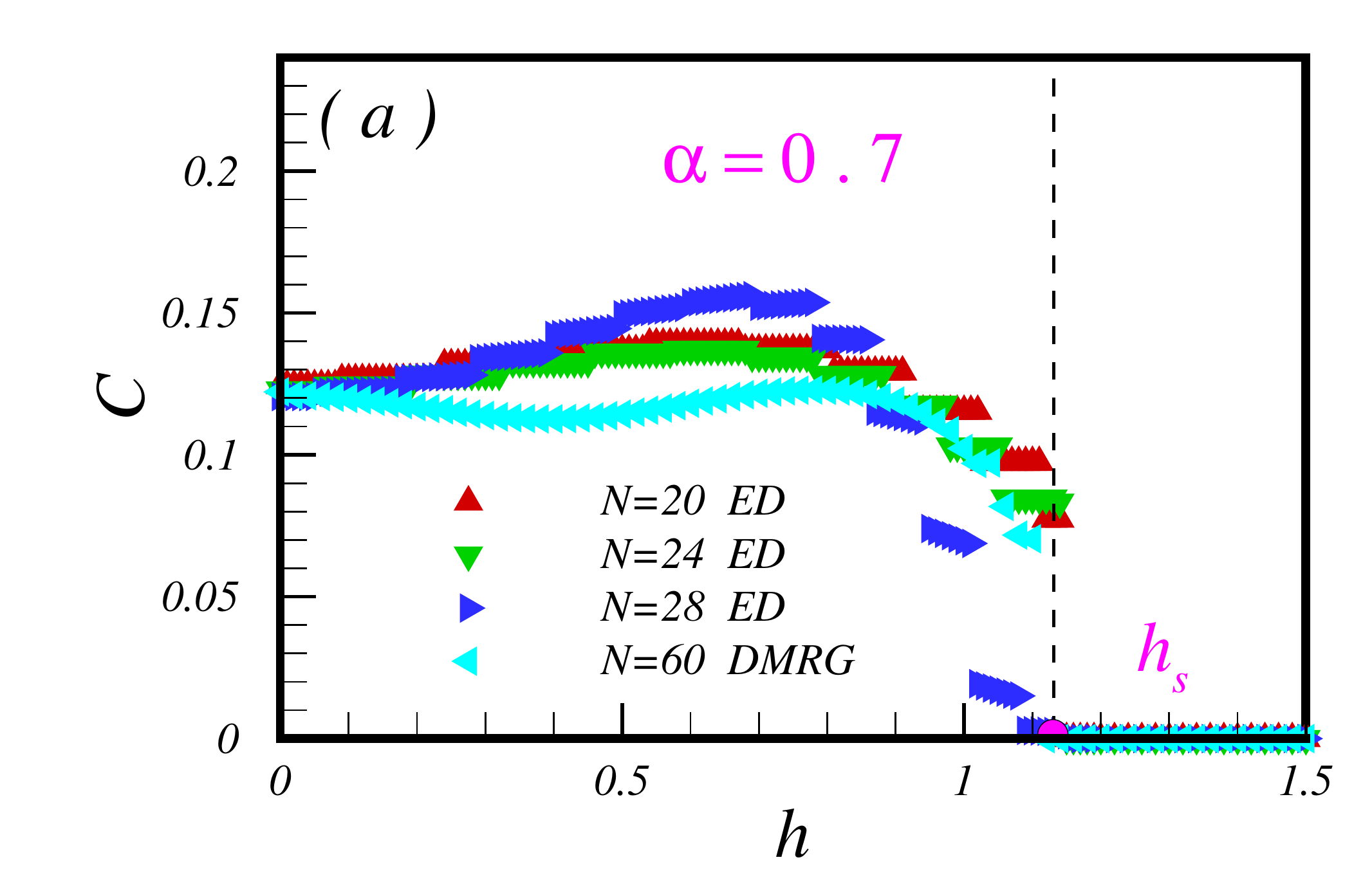,width=1.8in} \psfig{file=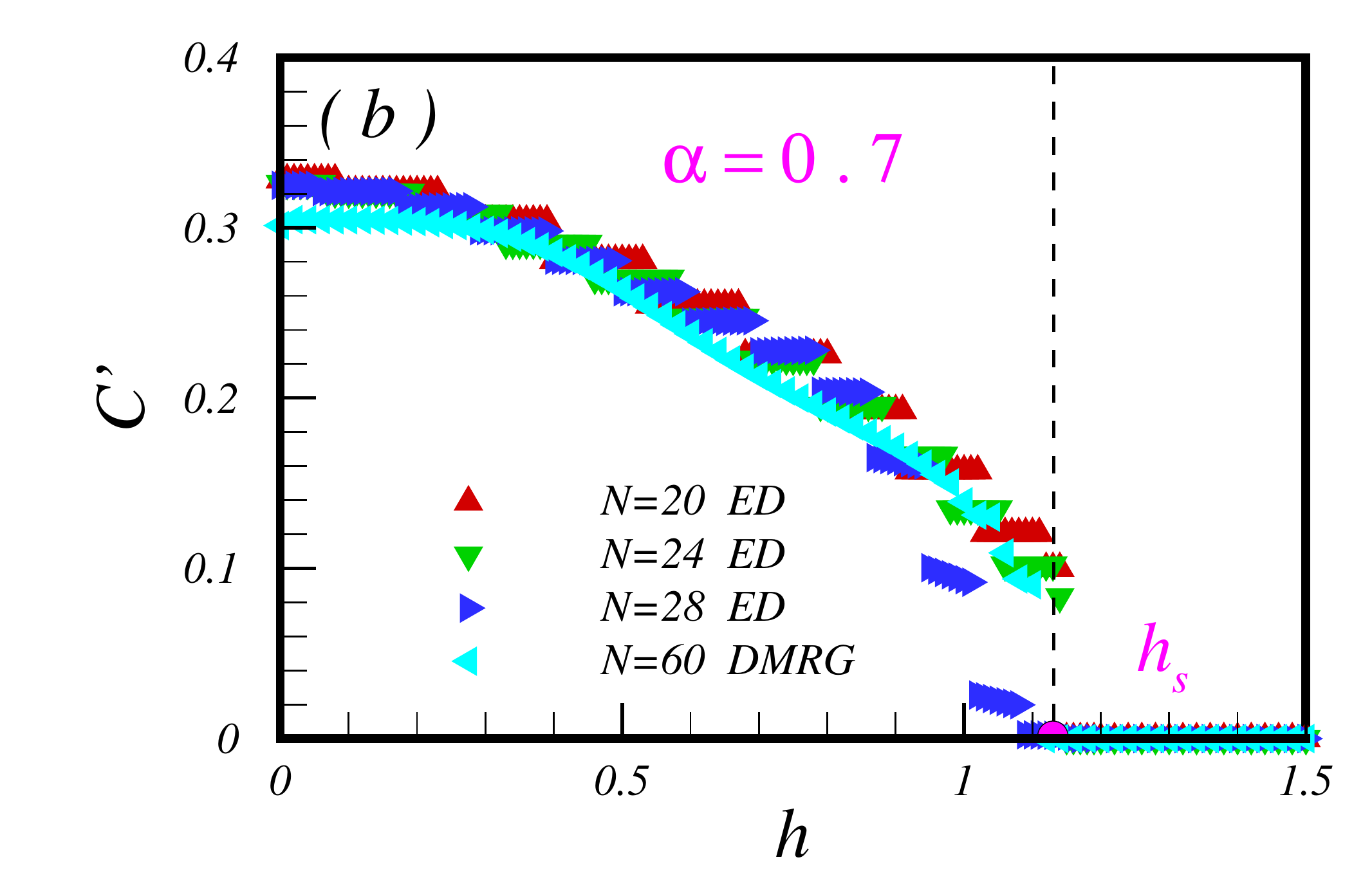,width=1.8in}}
\centerline{\psfig{file=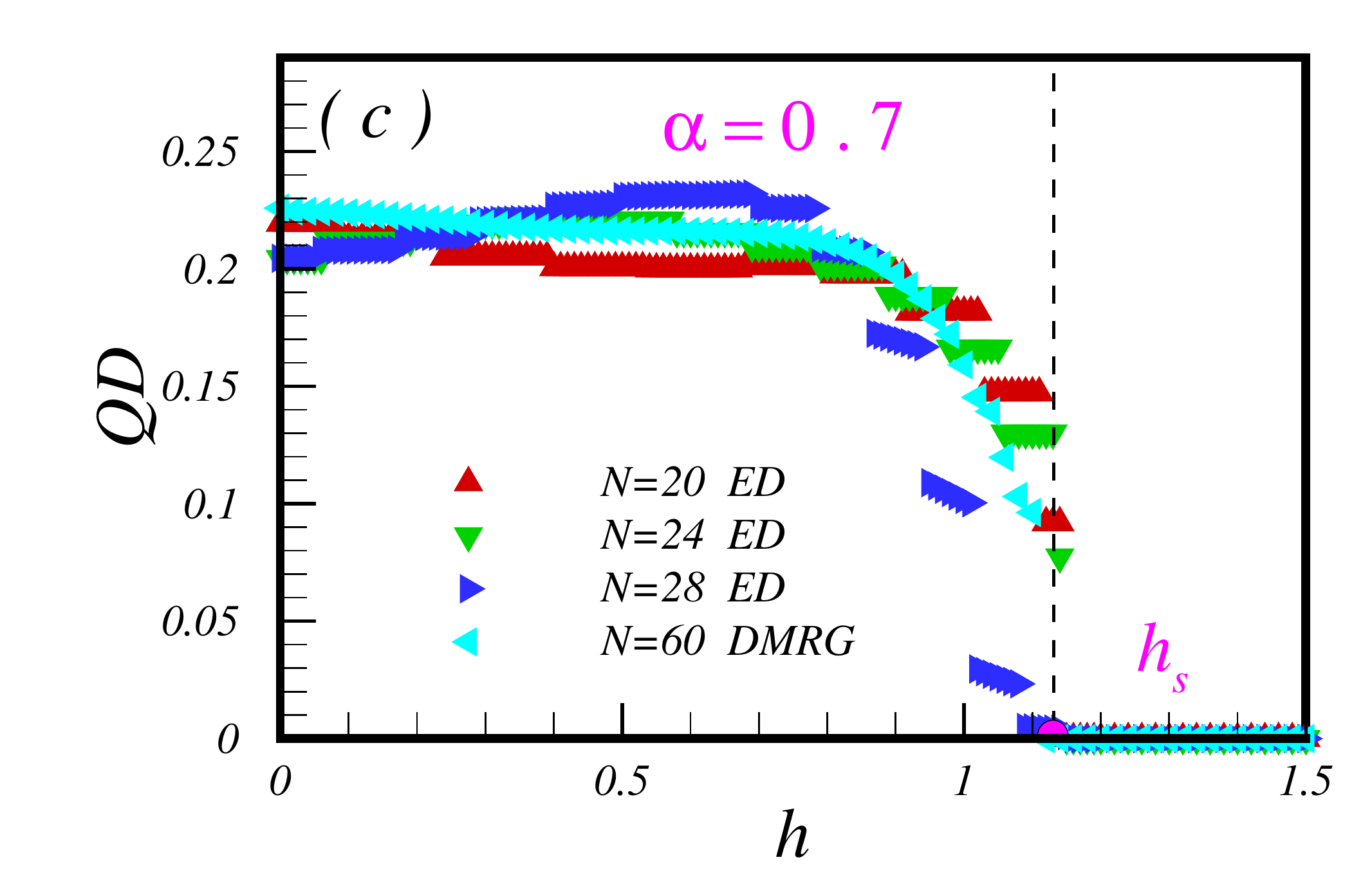,width=1.8in} \psfig{file=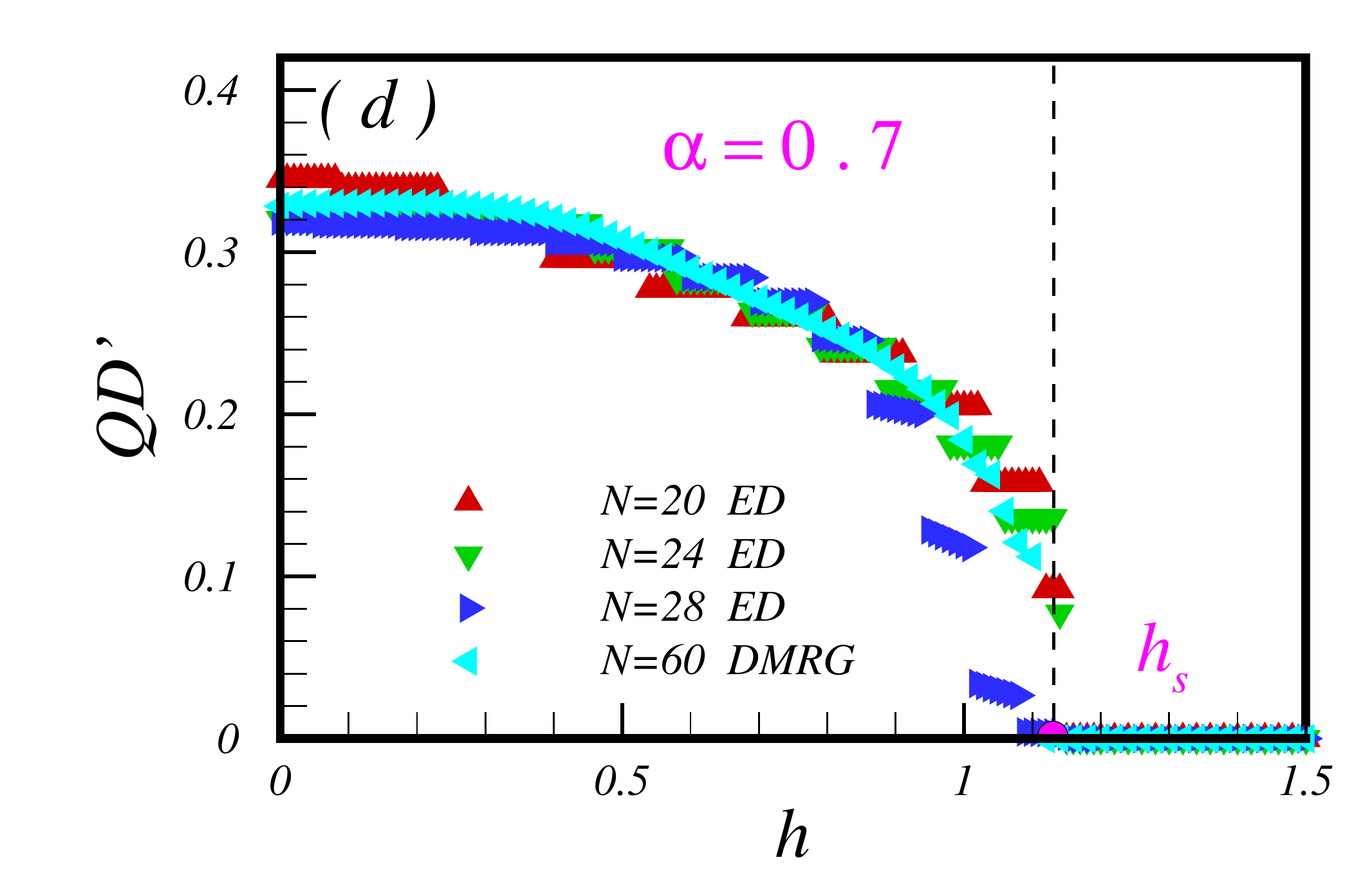,width=1.8in}}
\caption{(color online) The concurrence and the QD between pair of spins on bonds with exchange coupling ${J}$ ((a) and (c)) and ${J'}$ ((b) and (d)) versus the TF. Lanczos results are presented for $\alpha=0.2$ and clusters with $N=20, 24, 28$ spins and also DMRG results for $N=60$.   }
\label{Ch}
\end{figure}
\par
 Now lets us consider the transverse magnetic field, and probe the entanglement and QD evolution throughout the model. To this end, we fix the parameter $\alpha$ such as the model exists (i) at the N\'eel phase with $\alpha>\alpha_c$, (ii)  at dimerized phase with $\alpha<\alpha_c$.

Results for the case (i) with $\alpha=0.7$  are plotted in Fig.~\ref{Ch}. At $h=0$, as identified before, concurrence  is shared between all nearest-neighbor pair spins. By tuning the magnetic field, ${C}$ shows almost increasing behaviour  until the quantum critical region (see Figs.~\ref{Ch} (a)). As soon as the system enters to the quantum critical region,  ${C}$ decreases monotonically till disappearing at saturation filed $h_s$. That is expected at $h_s(\alpha)\simeq1.14$, as all of the spins are aligned in the direction of the field. On the other hand, as soon as the TF  turns on, ${C'}$ decreases and will be disappeared at the saturation TF, $h_s$ (see Figs.~\ref{Ch} (b)). No signature of the quantum critical region is seen in the behaviour of ${C'}$. It should be noted that the same behaviour as the $C$ and $C'$ is observe for the QD between pair of spins on bonds with exchange coupling $J$ (Figs.~\ref{Ch} (c)) and $J'$ (Fig.~\ref{Ch} (d)). Observed oscillations of the   quantum correlations result from the level crossing between the ground and the excited states of the model.

\begin{figure}[t]
\centerline{\psfig{file=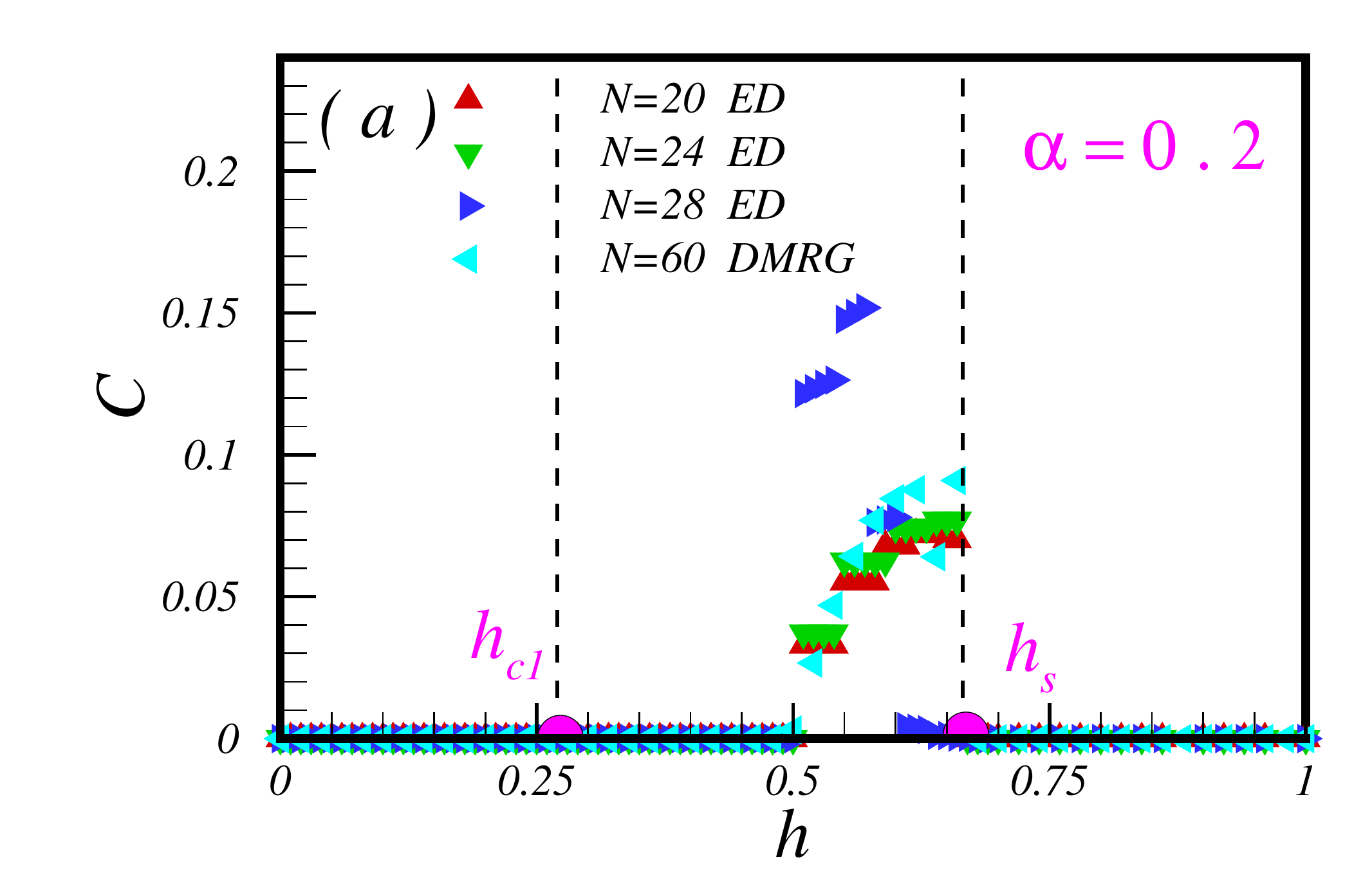,width=1.8in} \psfig{file=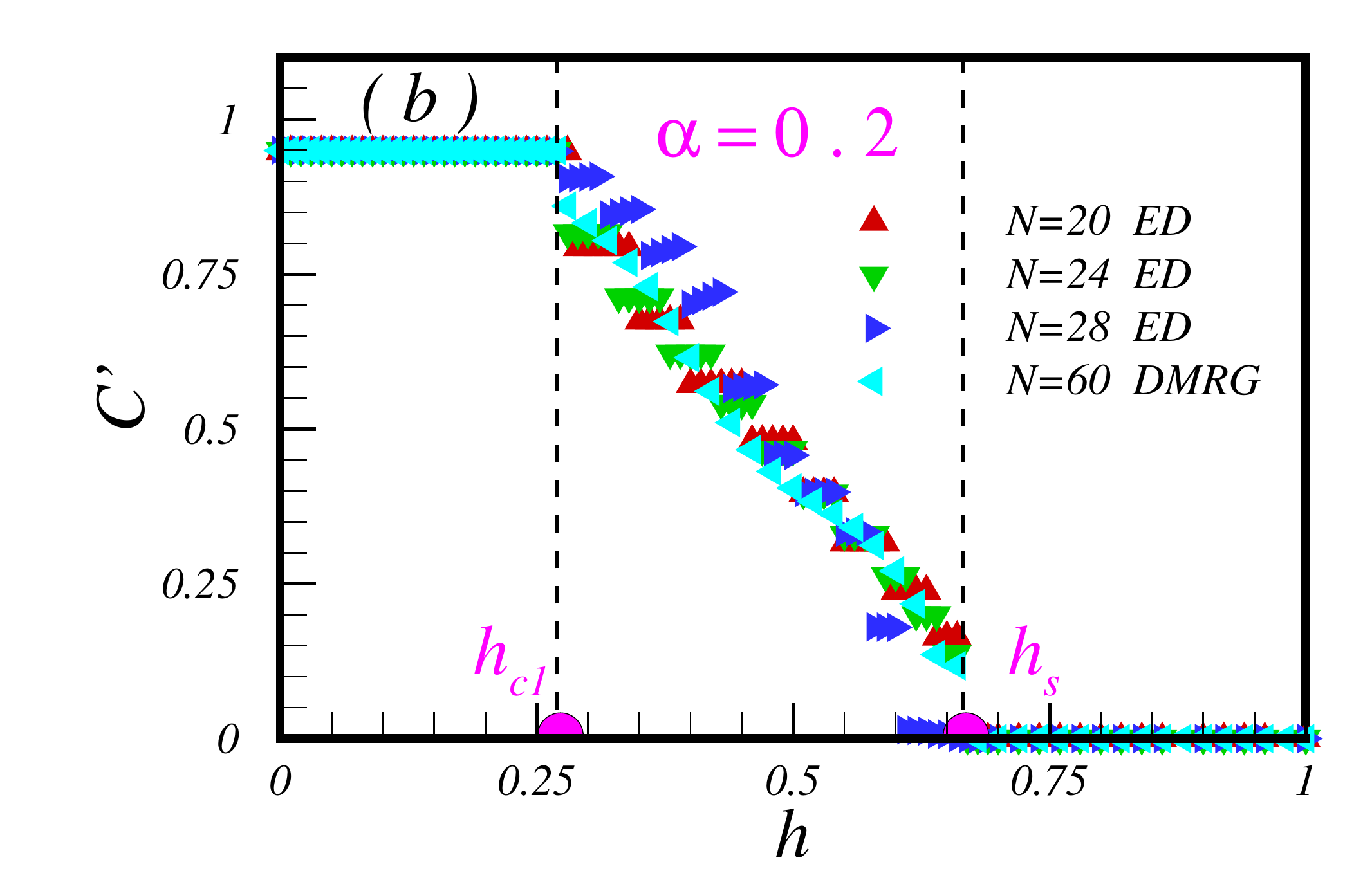,width=1.8in}}
\centerline{\psfig{file=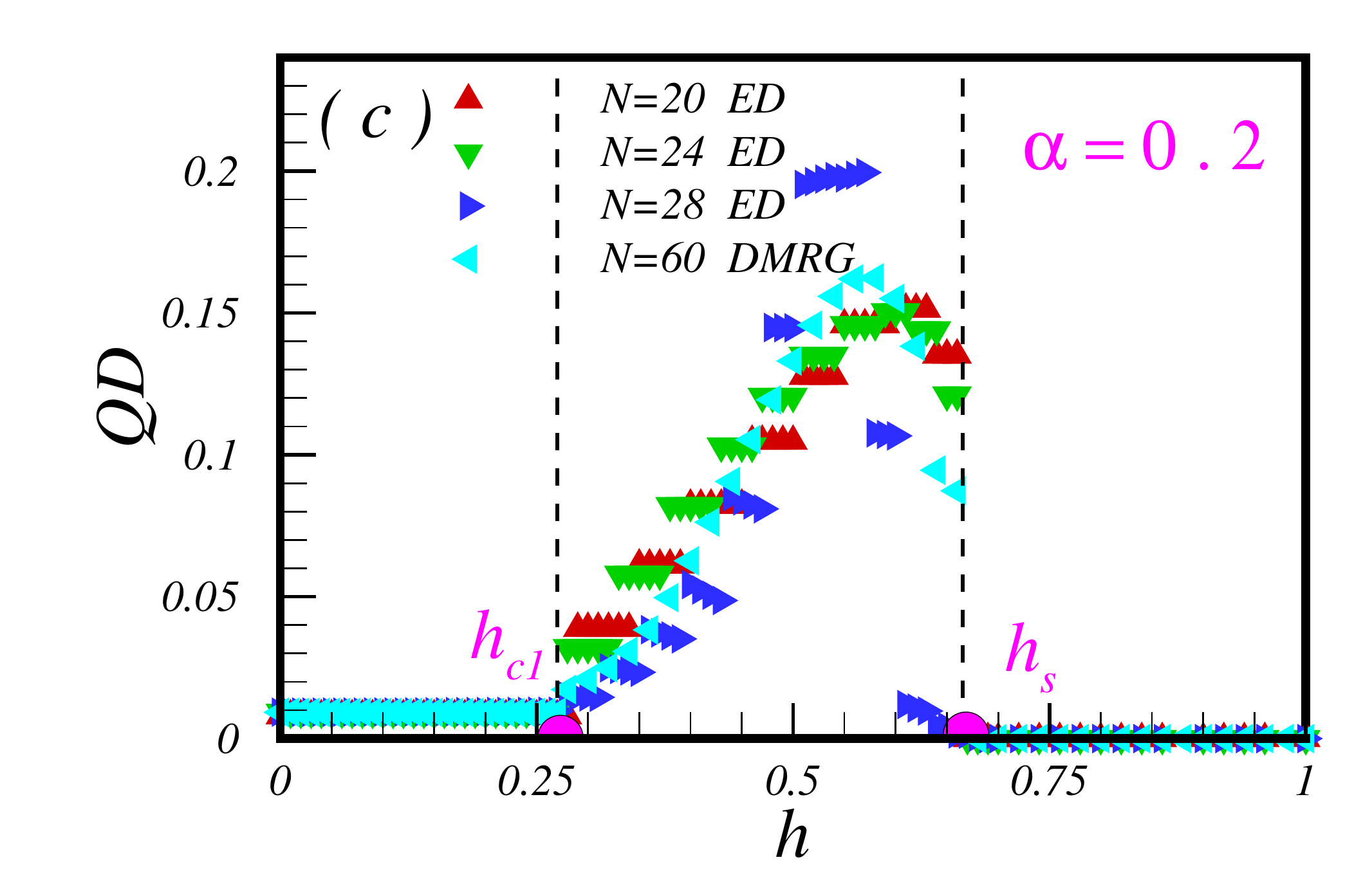,width=1.8in} \psfig{file=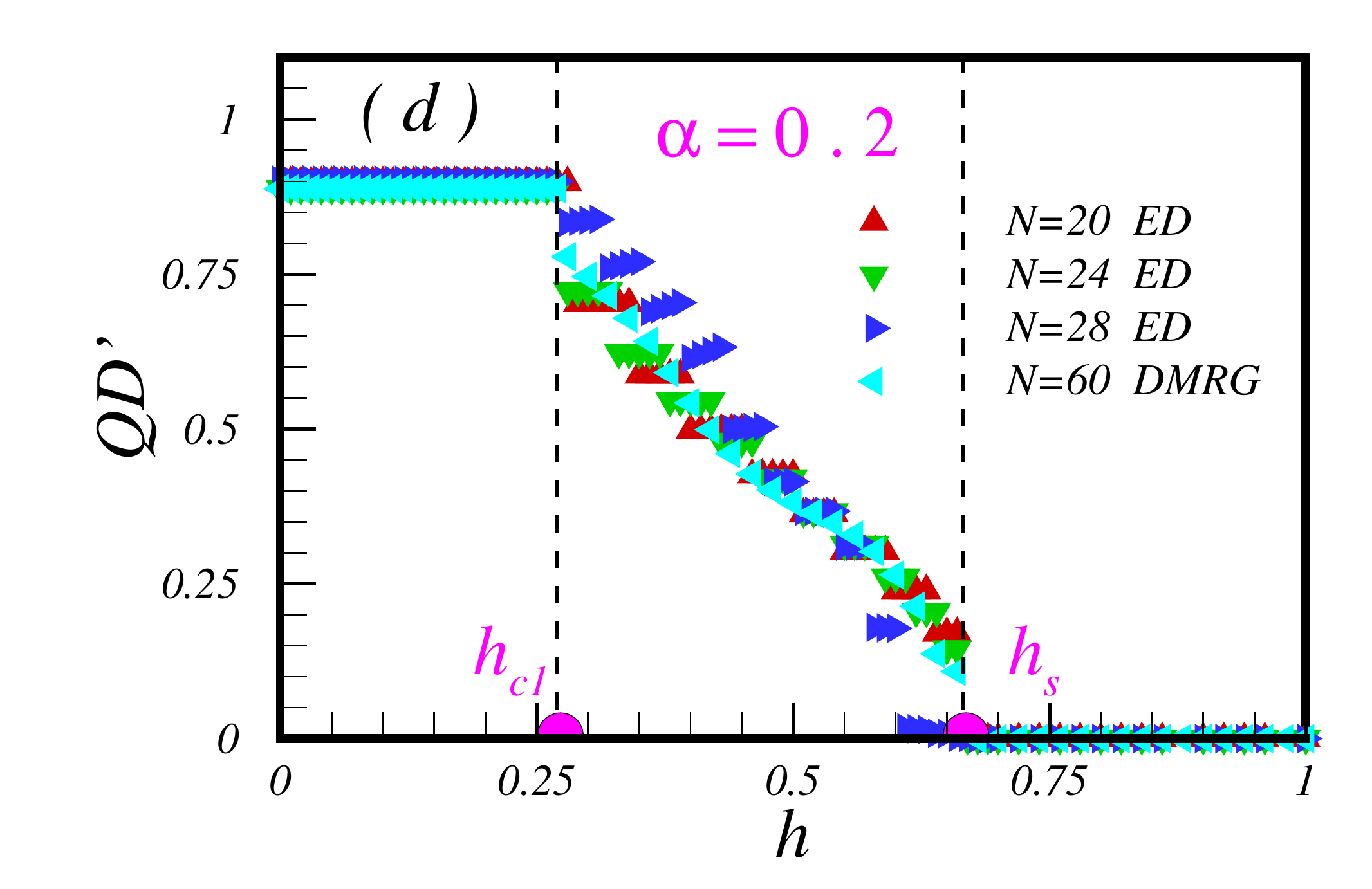,width=1.8in}}
\caption{(color online) The concurrence and the QD between pair of spins on bonds with exchange coupling $J$ ((a) and (c)) and $J'$ ((b) and (d)) versus the TF. Lanczos results are presented for $\alpha=0.2$ and clusters with $N=20, 24, 28$ spins and also DMRG results for $N=60$.   }
\label{Qh}
\end{figure}

We have done the same numerical experiment  when the system exists deep in  the dimerized phase with $\alpha=0.2$ and results are presented in Figs.~\ref{Qh}. Within this parameter, the model can be effectively assumed as an ensemble of singlet pairs that are weakly interacting.  It is known that by applying a TF, system remains in the gaped dimerized phase up to the first critical field $h_{c_{1}} (\alpha)$. With more increasing the field, the system goes to a canted N\'eel phase, and finally, at a saturation field $h_s (\alpha)$ becomes polarized. 

Interesting behaviour is seen in the results of the concurrence between pair of spins on bonds with exchange coupling ${J}$ (Figs.~\ref{Qh}-(a)). Despite  these pair of spins are not entangled in the absence of the TF, they remain unentangled in the canted N\'eel phase, which shows that low excited states of the pure dimerized model in the region $\alpha<\alpha_c$ are not entangled by considering $C$. More interesting, we have found the "magnetic entanglement" by increasing the TF.  The TF creates entanglement between pair of spins on bonds with exchange coupling $J$, at field $h=h_E (\alpha)$ and in the region $h_E(\alpha)<h<h_s$ an entangled region is observed. This entangled region shows that the high excited states of the pure dimerized model in the region $\alpha<\alpha_c$ are entangled by considering $C$.

As can be  seen from Figs.~\ref{Qh}-(b) and (d), in absence of TF, pair of spins on bonds with exchange coupling ${J'}$ are quantum correlated. These bonds remain entangled or quantum correlated with a constant value up to the first critical $h_{c_1}\simeq0.27$. By increasing  the field, $C'$ and $QD'$ develop a series of plateaus with a decreasing trend and vanish at the saturated field $h_s\simeq0.67$.  

Finally as is seen in Figs.~\ref{Qh}-(c), QD between  pair of spins on bonds with exchange coupling ${J}$, shows an almost zero-plateau up to first critical TF and exactly a zero-plateau in the region $h>h_s$. It is observed that as soon as the system enters into the canted  N\'eel phase, QD increases up to the vicinity of the saturation field $h_s$. Exactly at the saturation TF, QD will be zero and no quantum correlations is observed in the PM region. 

\section{CONCLUSION}\label{sec6 }

We considered a dimerized spin-1/2 XX honeycomb model in the presence of a transverse  magnetic field. At zero temperature the ground state phase diagram is known. In the absence of the field, there is a critical dimerization point $\alpha_c$, which separates the commensurate N\'eel and incommensurate dimer phases. In presence of the field, system becomes polarized at a saturationf ield, $h=h_s(\alpha)$. By placing the model at dimerized phase and changing the field, model undergoes a quantum phase transition from the dimer into the canted N\'eel phase at $h_{c_1} (\alpha)$. With a more increasing field, spins finally get aligned with the field at the saturation point $h_s$.
 
Equipped with the knowledge above, we tried to understand the entanglement feature of the model. To this end, we borrowed  concurrence and quantum discord (QD) observable from the quantum information context.  We focused on the quantum correlations among the nearest-neighbour pair of spins on finite clusters using the complimentary numerical Lanczos and DMRG techniques. Critical dimerization point, $\alpha_c$, is obtained from the concurrence. In presence of the field, we observed the "magnetic entanglement" region between $h_E(\alpha)<h<h_s(\alpha)$, which an entanglement creates between pairs of unentangled spins.  Exploiting quantum entanglement features to study exotic magnetic phases at zero temperature has privileges compared to the  Landau theory, as the definition of a proper order parameter is not easy. This work could potentially be extended to check the resonating valence bonds (RVB) state or quantum spin liquid (QSL) phase in the honeycomb lattice\cite{R23,R24}.

\section{Acknowledgement}  
J. V gratefully acknowledge support from Deutsche Forschungsgemeinschaft (DFG) KE-807/22-1.

\bibliography{Ref/ref}

\end{document}